\documentclass[twocolumn,showpacs,preprintnumbers,amsmath,amssymb]{revtex4}

\usepackage{graphicx}% Include figure files
\usepackage{dcolumn}% Align table columns on decimal point
\usepackage{bm}% bold math

\begin{document}
\title{General Dynamics of Topology and Traffic on Weighted Technological Networks}
\author{Wen-Xu Wang$^{1}$, Bo Hu$^{1}$}
\author{Gang Yan$^{2}$}
\author{Qing Ou$^{1}$}
\author{Bing-Hong Wang$^{1}$}
\email{bhwang@ustc.edu.cn, Fax:+86-551-3603574.}
\affiliation{%
$^{1}$Nonlinear Science Center and Department of Modern Physics,
$^{2}$Department of Electronic Science and Technology,\\
University of Science and Technology of China, Hefei, 230026, P.R.
China
}%

\date{\today}

\begin{abstract}
For most technical networks, the interplay of dynamics, traffic
and topology is assumed crucial to their evolution. In this paper,
we propose a traffic-driven evolution model of weighted
technological networks. By introducing a general strength-coupling
mechanism under which the traffic and topology mutually interact,
the model gives power-law distributions of degree, weight and
strength, as confirmed in many real networks. Particularly,
depending on a parameter $W$ that controls the total weight growth
of the system, the nontrivial clustering coefficient $C$, degree
assortativity coefficient $r$ and degree-strength correlation are
all in consistence with empirical evidences.
\end{abstract}


\maketitle

The past few years have witnessed a great devotion of physicists
to understand and characterize the underlying mechanisms of
complex networks including the Internet \cite{Internet}, the WWW
\cite{WWW}, the scientific collaboration networks (SCN)
\cite{CN1,CN2} and world-wide airport networks
(WAN)\cite{air1,air2}. So far, researches on networks have mainly
focused on unweighted graphs. Barab\'asi and Albert have proposed
a well-known model (BA model) that introduces the degree
preferential attachment mechanism to mimic unweighted growing
networks \cite{BA}. Most recently, the availability of more
complete empirical data has allowed scientists to consider the
variation of the weights of links that reflect the physical
characteristics of many real networks. Obviously, there is a need
for a modelling approach to complex networks that goes beyond the
purely topological point of view. Alain Barrat, et al. presented a
model (BBV model) that integrates the topology and weight
dynamical evolution to study the growth of weighted networks
\cite{BBV}. Their model yields scale-free properties of the
degree, weight and strength distributions, controlled by an
introduced parameter $\delta$. However, its weight dynamical
evolution is triggered only by newly added vertices, resulting in
few satisfying interpretions to the collaboration networks or the
airport systems. In fact, the dynamics and properties of social
and technological networks are quite different and should be
addressed individually. It is well-known that networks are not
only specified by their topology but also by the dynamics of
weight (e.g. information flow) taking place along the links. For
instance, the heterogeneity in the intensity of connections may be
very important in understanding technological systems. The amount
of traffic characterizing the connections of communication systems
or large transport infrastructure is fundamental for a full
description of these networks \cite{top10}. Take the WAN for
example: each given edge weight $w_{ij}$ (traffic) is the number
of available seats on direct flight connections between the
airports $i$ and $j$. Weighted networks are often described by an
adjacency matrix $w_{ij}$ which represents the weight on the edge
connecting vertices $i$ and $j$, with $i,j=1,\ldots,N$, where $N$
is the size of the network. We will only consider undirected
graphs, where the weights are symmetric ($w_{ij}=w_{ji}$). As
confirmed by measurements, complex networks often exhibit a
scale-free degree distribution $P(k)\thicksim k^{-\gamma}$ with
2$\leq\gamma\leq$3 \cite{air1,air2}. The weight distribution
$P(w)$ that any given edge has weight $w$ is another significant
characterization of weighted networks, and it is found to be heavy
tailed, spanning several orders of magnitude \cite{ref1}. A
natural generalization of connectivity in the case of weighted
networks is the vertex strength described as
$s_{i}=\sum_{j\in\Gamma(i)}w_{ij}$, where the sum runs over the
set $\Gamma(i)$ of neighbors of node $i$. The strength of a vertex
integrates the information about its connectivity and the weights
of its links. For instance, the strength in WAN provides the
actual traffic going through a vertex and is obvious measure of
the size and importance of each airport. Empirical evidence
indicates that in most cases the strength distribution has a fat
tail \cite{air2}, similar to the power law of degree distribution.
Highly correlated with the degree, the strength usually displays
scale-free property $s\thicksim k^{\beta}$ \cite{traffic-driven,
empirical}.

The previous models of complex networks always incorporate the
(degree or strength) preferential attachment mechanism, which may
result in scale-free properties. Essentially, this mechanism just
describes interactions between the newly-added node and the old
ones. Actually, such interactions also exist between old nodes.
Perhaps, the most reasonable and simplest way to express such
interactions is by the product form of related vertex strengths,
i.e. the pairwise interaction between vertices $i$ and $j$ is
proportional to $s_is_j$ (strength coupling form). Let's review BA
model: a new vertex $n$ is added with $m$ edges that are randomly
attached to an existing vertex $i$ according to the degree
preferential probability, which can be written in the product form
of degrees
\begin{equation}
\Pi_{n\rightarrow
i}^{BA}=\frac{k_i}{\sum_{j}k_j}=\frac{k_nk_i}{\sum_{j}k_nk_j}.
\end{equation}
Analogously, in BBV networks one can rewrite the strength
preferential probability:
\begin{equation}
\Pi_{n\rightarrow
i}=\frac{s_i}{\sum_{j}s_j}=\frac{s_ns_i}{\sum_{j}s_ns_j}.
\end{equation}
We argue that such interactions (actually driven by traffic) exist
between old vertices in the same way, and will considerably affect
the flows between them: First, new edges should be allowed to add
between old nodes; second, the pre-existing traffic flows along
the links will be updated with the growth of networks. Indeed,
physical interaction of nodes plays a crucial role in determining
the network topology during its dynamical evolution. Our above
perspectives have been partly inspired by the work of Dorogovtsev
and Mendes (DM) \cite{DM} who proposed a class of undirected and
unweighted models where new edges are added between old sites
(internal edges) and existing edges can be removed (edge removal).

In the letter, we present a model for weighted technological
networks that considers the topological evolution under the
general traffic-driven interactions of vertices. It can mimic the
reinforcement of internal connections and the evolution of many
infrastructure networks. The diversity of scale-free
characteristics, nontrivial clustering coefficient, assortativity
coefficient and strength-degree correlation that have been
empirically observed can be well explained by our microscopic
mechanisms. Moreover, in contrast with previous models where
weights are assigned statically \cite{ref2,ref3} or rearranged
locally (BBV model), we allow the flows to be widely updated.

The model starts from an initial configuration of $N_{0}$ vertices
connected by links with assigned weight $w_{0}$. The model is
defined on two coupled mechanisms: the topological growth and the
strengths' dynamics:

(i){\it Strengths' Dynamics}. From the beginning of the evolution,
all the possible (existing or not) connections at each time step
are supposed to update their weights according to the
strength-coupling mechanism:
\begin{equation}
 w_{ij}\rightarrow\left\{
    \begin{array}{cc}
        w_{ij}+1, &\mbox{with probability $Wp_{ij}$}\\
        w_{ij}, &\mbox{with probability $1-Wp_{ij}$,}
    \end{array}
    \right.
\end{equation}where\begin{equation}
p_{ij}=\frac{s_is_j}{\sum_{a<b}s_as_b},
\end{equation}integrates the strength coupling
of vertices $i$ and $j$, and determines the increment probability
of weight $w_{ij}$ (if $i$ and $j$ are unconnected, $w_{ij}=0$).
The total weight of the edges in statistical sense is modified by
the amount $<\sum_{i<j} \Delta w_{ij}>=W$, which is assumed
constant for simplicity. This parameter reflects the growing speed
of the network's total traffic load; for instance, the increasing
rate of total information flow in a communication system. The
always growing traffic plays the driving role in network
evolution. One may notice that $Wp_{ij}$ is very likely to exceed
one if the initial number of nodes $N_0$ is small. When $Wp_{ij}$
exceeds one, it is automatically assumed to be one. This treatment
of $Wp_{ij}$ will probably affect the initial network evolution,
while it is not significant for discussing the statistical
measures, as they are almost independent of initial states.

(ii){\it Topological Growth.} At the same time step, a new vertex
$n$ is then added with $m$ edges that are randomly attached to an
existing vertex $i$ according to the strength preferential
probability $\Pi_{n\rightarrow i}$ (Eq. (2)). The weight of each
new edge is also fixed to $w_{0}$. In fact, the strength
preferential attachemnt is essentially the same with the
 mechanism traffic-driven interactions we have argued.

\begin{figure}
\scalebox{0.25}[0.25]{\includegraphics{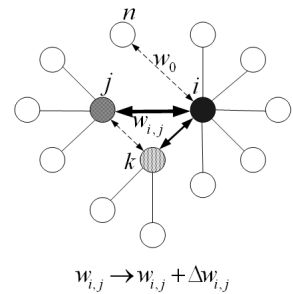}}
\caption{\label{fig:epsart} Illustration of the evolution
dynamics. A new node $n$ connects to a node $i$ with probability
proportional to $s_i/\sum_j s_j$. The thickness of nodes and links
respectively represents the magnitude of strength and weight. New
connections (dashed) can be built between pre-existing nodes and
the bilateral links represent the traffic growing process along
links under the general mechanism of strength couplings.}
\end{figure}

The network provides the substrate on which numerous dynamical
processes occur. In previous models, traffic was often assumed
just as an appendix to the network structure. Actually, traffic
and the underlying topology are mutually correlated and it is very
important to define appropriate quantities and measures capable of
capturing how all these ingredients participate in the formation
of complex networks \cite{top10}. Technology networks provide
large empirical database that simultaneously captures the topology
and the dynamics taking place on it. For Internet, the information
traffic between routers (nodes) can be represented by the
corresponding edge weight. The total traffic that each router
deals with can be denoted by the node strength, which also
represents the importance of given router. The increasing
information flow as an internal demand always spurs the expansion
of technological networks. Specifically, the largest contribution
to the growth is given by the emergence of links between already
existing nodes. This clearly points out that the Internet growth
is strongly driven by the need of a redundancy wiring and an
increasing need of available bandwidth for data transmission
\cite{empirical}. On one end, newly-built links (between existing
routers) are supposed to preferentially connect high strength
routers, because otherwise, it would lead to the unnecessary
traffic congestion along indirect paths that connect those high
strength nodes. Naturally, traffic along existing links between
high strength routers, in general, increases faster than that
between low strength routers. All the points are reflected in our
strength-coupling mechanism. On the other end, new routers
preferentially connect to routers with larger bandwidth and
traffic handling capabilities (the strength driven attachment).
Those phenomena also exist in airport system, power grid, and
railroad network; and they could be explained by the
traffic-driven mechanism of our model. For power grid and railroad
network, the cost by distance has a distinct effect to their
topological properties. Their degree distributions, for example,
are not scale-free. In a word, topology and traffic interact with
each other in networks under general interactions of vertices
driven by the internal increasing traffic demand.

The model time is measured with respect to the number of nodes
added to the graph, i.e. $t=N-N_0$, and the natural time scale of
the model dynamics is the network size $N$. In response to the
demand of increasing traffic, the system must expand. With a
certain size, one technological network assumably has a certain
ability to handle certain traffic load. Therefore, it could be
reasonable to suppose that the total weight on the networks
increases synchronously by the natural time scale. That is why we
assume $W$ as a constant. This assumption also bring us the
convenience of analytical discussion. Using the continuous
approximation, we can treat $k$, $w$, $s$ and the time $t$ as
continuous variables \cite{Internet, BA}. Then Eq. (3) indicates:
\begin{equation}
\frac{dw_{ij}}{dt}=\frac{2Ws_is_j}{\sum_{a,b(a\neq
b)}s_as_b}=\frac{2Ws_is_j}{\sum_as_a\sum_{b(\neq a)}s_b}.
\end{equation}
The strength $s_i$ of vertex $i$ can increase either if a
new-added node connects to $i$ by the topological growth dynamics
or any possible (exsiting or not) connections to $i$ are updated
by the strengths' dynamics:
\begin{eqnarray}
\frac{ds_i}{dt}=\frac{\sum_{j(\neq
i)}2Ws_is_j}{\sum_as_a\sum_{b(\neq
a)}s_b}+\frac{ms_i}{\sum_ls_l}=\frac{2W+m}{2W+2m}\frac{s_i}{t}
\end{eqnarray}
where the last expressions recovered by noticing that
$\sum_is_i(t)\approx2(m+W)t$. From Eq. (6), one can analytically
obtain the power-law distribution of strength ($P(s)\sim
s^{-\alpha}$) with the exponent \cite{BA, BBV}:
$\alpha=2+m/(m+2W)$. Obviously, when $W=0$ the model is
topologically equivalent to the BA network and the value
$\alpha=3$ is recovered. For larger values of $W$, the
distribution is gradually broader with $\alpha\rightarrow2$ when
$W\rightarrow\infty$.

\begin{figure}
\scalebox{0.70}[0.60]{\includegraphics{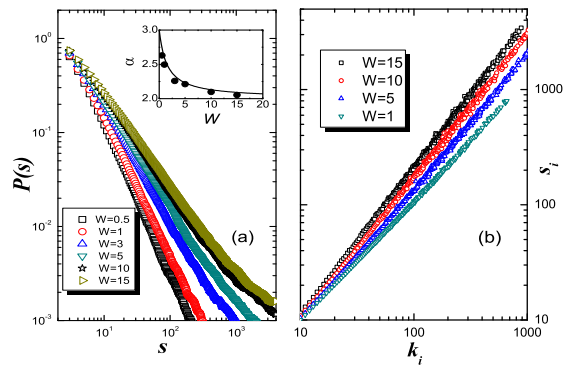}}
\caption{\label{fig:epsart} (a) Probability distribution $P(s)$.
Data are consistent with a power-law behavior $s^{-\alpha}$. In
the inset we give the value of $\alpha$ obtained by data fitting
(filled circles), together with the analytical expression
$\alpha=2+m/(m+2W)$(line). The data are averaged over 20 networks
of size N=5000. (b) Strength $s_i$ versus $k_i$ for different $W$
(log-log scale). Linear data fitting gives slope 1.04, 1.17, 1.25
and 1.30 (from bottom to top), demonstrating the correlation of
$s\sim k^{\beta}$.}
\end{figure}

\begin{figure}
\scalebox{0.70}[0.60]{\includegraphics{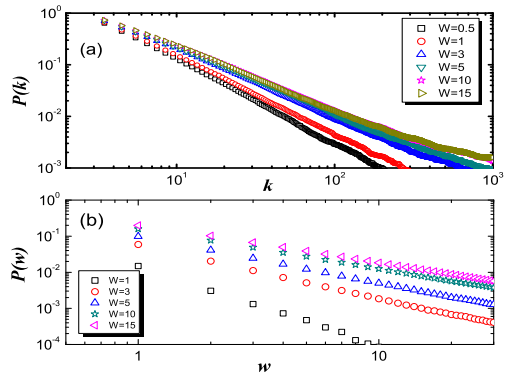}}
\caption{\label{fig:epsart} (a) Probability distribution of the
degrees $P(k)\sim k^{-\gamma}$. (b) Probability distribution of
the weights $P(w)\sim w^{-\eta}$. The data are averaged over 20
networks of size N=5000.}
\end{figure}

We performed numerical simulations of networks generated by
choosing different values of $W$ and fixing $N_{0}=3$, $m=3$ and
$w_{0}=1$. We have checked that the scale-free properties of our
model networks are almost independent of the initial conditions.
Numerical simulations are consistent with our theoretical
predictions, verifying again the reliability of our present
results. Fig. 2(a) gives the probability distribution $P(s)\sim
s^{\alpha}$, which is in good agreement with the theoretical
predictions. We also report the average strength $s_i$ of vertices
with degree $k_i$, which displays a nontrivial power-law behavior
$s\sim k^{\beta}$ as confirmed by empirical measurement. Unlike
BBV networks (where $\beta=1$), the exponent $\beta$ here varies
with the parameter $W$ in a nontrivial way as shown in Fig. 2(b).
Moreover, the major difference between our model and DM network is
reflected in the strength-degree correlation graph. Though DM
model allows the emergence of internal edges, it could not mimic
the reinforcement of pre-existing connections in that it is
unweighted. The nontrivial $s\sim k^{\beta}$ correlation
demonstrates the significant part of weight increment along
existing edges, and thus implies that our model is reasonable in
the light of traffic flow. More importantly, one could check the
scale-free property of degree distribution ($P(k)\sim
k^{-\gamma}$) by combining $s\sim k^{\beta}$ with $P(s)\sim
s^{-\alpha}$. Considering $P(k)dk=P(s)ds$, the exponent $\gamma$
is easily calculated: $\gamma=\beta(\alpha-1)+1$. The scale-free
properties of weight and degree obtained from simulations are
presented in Fig. 3(a)(b). Finally, it is worth remarking that for
the BA networks, the clustering coefficient is nearly zero, far
from the practical nets that exhibit a variety of small-world
properties. In the present model, however, clustering coefficient
$C$ is found to be a function of $W$ (Fig. 4(a)), also supported
by empirical data of a broad range.

\begin{figure}
\scalebox{0.70}[0.55]{\includegraphics{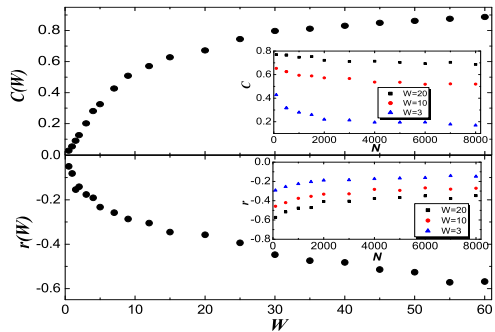}}
\caption{\label{fig:epsart} (a) Clustering coefficient $C$
depending on the parameter $W$. In the inset, we report the
evolution of clustering coefficient (or $C$ versus $N$) which
converges soon. (b) Degree-degree correlation $r$ depending on
$W$. In the inset, we report its evolution which converges soon.}
\end{figure}

In the social networks, connections between people may be
assortative by language or by race. Mixing can also be
disassortative, i.e. vertices in the network preferentially form
connections to others unlike them. Newman proposed some simple
measures for these types of mixing, which we call assortativity
coefficients \cite{mixing}. In the case of mixing by vertex
degree, a remarkable pattern emerges. Almost all the social
networks studied show positive assortativity coefficients while
all others, including technological and biological networks, show
negative coefficients. It is not clear if this is a universal
property; the origin of this difference is not understood either.
In our views, it represents a feature that should be addressed in
each network individually. We argue that the adaptive evolution of
topology in response to the increasing traffic is the major cause
of disassortative mixing of technological networks. Using the
formula defined in Ref. \cite{mixing} (Eq. (26)), we calculate the
degree assortativity coefficient (or degree-degree correlation)
$r$ of the graphs generated by our model. Simulations given in the
Fig. 4(b) are supported by empirical measurements \cite{mixing}.
The restriction of our model to technological networks is for that
there are few empirical data for statistical analysis on
``weighted" biological networks, where many interacting mechanisms
are far from present knowledge either. But hopefully, our model
may be very beneficial for future understanding or characterizing
biological networks and social ones, as it generates many
topological properties observed in those real networks. Due to its
apparent simplicity and the variety of controllable results, we
believe that some of its extensions will probably help address the
other two classes of networks.

In conclusion, the universal interactions of nodes and internal
traffic demand of the system will determine the topology evolution
of technological network. This general traffic-driven mechanism
provides a wide variety of scale-free behaviors, clustering
coefficient and nontrivial correlations, depending on the
parameter $W$ that governs the total weight growth. All the
results are supported by empirical data. Therefore, our present
model for all practical purposes will demonstrate its applications
in future weighted network research.

We gratefully thank Yan-Bo Xie and Tao Zhou for useful discussion.
This work is funded by NNSFC under Grant No. 10472116 and No.
70271070.

\end{document}